\documentclass[prl,superscriptaddress,amsmath,amssymb,floatfix,twocolumn,amsfonts]{revtex4-2}

\usepackage{times}
\usepackage[varg]{txfonts}
\usepackage{textcomp}
\usepackage{graphicx}
\usepackage{subfigure}
\usepackage{subeqnarray}
\usepackage{mathtools}
\usepackage{mathrsfs}
\usepackage{color}
\usepackage[colorlinks=true,citecolor=blue,urlcolor=blue,linkcolor=blue,hyperindex]{hyperref}
\usepackage{braket}
\usepackage{overpic}
\usepackage{ulem}
\usepackage[version=3]{mhchem}

\begin{document}

\title{Landau levels in the mixed state of two-dimensional nodal superconductors: models, featured magneto-optical response and quantized thermal Hall effect}

\author{Zhihai Liu}
\affiliation{College of Physics and Optoelectronic Engineering, Shenzhen University, Shenzhen 518060, China}
\author{Bin Wang}
\affiliation{College of Physics and Optoelectronic Engineering, Shenzhen University, Shenzhen 518060, China}
\author{Luyang Wang}
\email{wangly@szu.edu.cn}
\affiliation{College of Physics and Optoelectronic Engineering, Shenzhen University, Shenzhen 518060, China}

\date{\today}

\begin{abstract}
Landau quantization of low-energy quasiparticles (QPs) in the mixed state of gapless superconductors is a celebrated problem. So far, the only superconducting system that has been shown to host Landau levels (LLs) of Bogoliubov QPs is the Weyl superconductor. Here, we first investigate the QPs in the mixed state of two Weyl superconductors, an intrinsic one and a heterostructure one, and reveal that the QP states in the former do not form LLs, in contrast to those in the latter where LLs of QPs are shown to exist. The key is whether the low-energy Hamiltonian respects a generalized chiral symmetry. Following the analysis, we show that a two-dimensional superconducting system -- a topological insulator-superconductor heterostructure respecting the generalized chiral symmetry -- exhibits LLs with Chern number $\pm1$ in the mixed state. We also show featured responses resulting from the LLs, including peaked magneto-optical conductivity and quantized thermal Hall conductivity, which could be used as experimental probes to detect LLs in superconductors. 
\end{abstract}
 
\maketitle

\textit{Introduction.---}
In a strong magnetic field, Landau level (LL) quantization of electron states in two-dimensional (2D) systems can lead to distinct physical phenomena, such as the quantum Hall effect~\cite{Ezawa2008quantum}. However, such quantization of quasiparticles (QPs) has rarely been reported in (the mixed state of) superconductors. Previously, a Dirac-LL structure that scales with $\sqrt{nB}$ was predicted in the QP spectrum of the mixed state of $d$-wave superconductors~\cite{SchriefferPRL1998,Anderson.arxiv1998}. Nevertheless, by involving the effect of spatially varying supercurrents in the vortex lattice~\cite{Melnikov.JPCM1999}, Franz and Te\v{s}anovi\'{c} subsequently demonstrated Bloch waves instead of LLs in the QP spectrum~\cite{Franz.PRL2000}. Similar investigations have been performed for $s$-wave and $p$-wave superconductors~\cite{Vafek.PRB63.2001}.

The scenario in Weyl superconductors, which feature Bogoliubov-Weyl (BW) nodes within the superconducting gap, seems to be different. More recently, Pacholski \textit{et al.} have revealed a Dirac-LL of Bogoliubov particles in the mixed state of a heterostructure Weyl superconductor, which exhibits a topologically protected dispersionless zeroth LL~\cite{Pacholski.PRL2018}. Furthermore, a tilted BW cone in the heterostructure Weyl superconductor, which includes the type-II (overtilted) node that connects an electronlike and a holelike Bogoliubov Fermi pocket, can show squeezed LLs with reduced spacings~\cite{ZhihaiLiuPRB2024}, and the dispersionless zeroth LL is protected by a generalized chiral symmetry~\cite{KawarabayashiPRB2011,KawarabayashiIJMP2012}. These phenomena are quite similar to those observed in normal Weyl semimetals~\cite{TchoumakovPRL2016,YuZhi-MingPRL2016}.

In this work, we first reveal that the low-energy QPs in an intrinsic Weyl superconductor are different from those in a heterostructure Weyl superconductor. In the vortex lattice of the former, the supercurrent velocity (where the external magnetic vector potential enters) does not couple to the BW fermions and breaks the generalized chiral symmetry for tilted cones. In contrast, the supercurrent velocity in the mixed state of the latter always couples to the BW fermions as a vector potential, and the generalized chiral symmetry is respected. Consequently, LL quantization of Bogoliubov QPs is achieved in the heterostructure Weyl superconductor~\cite{Pacholski.PRL2018,ZhihaiLiuPRB2024}.

Following the symmetry principle, we propose a 2D superconducting system realized through a topological insulator-superconductor heterostructure, which may exhibit Dirac-LLs and graphene-like magneto-optical conductivity in the vortex state. Our calculations of the Chern number of each (non-degenerate) LL yield $\pm 1$, which is the same as those in the normal state, a layer of a topological insulator (TI)~\cite{BAndrei2013}. As a result, a quantized Hall response may be anticipated in the 2D system. In fact, in the low-temperature limit, we obtain a quantized transverse conductivity tensor $\tilde{\sigma}_{xy}$ and a quantized thermal Hall conductivity $\kappa_{xy}/T$. 

\textit{QPs in the mixed state of Weyl superconductors.---}
The heterostructure Weyl superconductors are engineered by alternately stacking $s$-wave superconductor and Weyl semimetal (WSM) (or TI ) layers~\cite{Meng.PRB2012,Meng.PRB2017,Nakai.PRB2020}. Conventional superconductivity is induced in the WSM layers by the superconducting proximity effect. The Weyl superconductivity in the heterostructure can be described by a Bogoliubov-de Gennes (BdG) Hamiltonian, in which each Weyl node of the WSM layers is doubled into an electron and hole node, and then mixed by the pair potential. In a magnetic field ${\bf B}=B_0\hat{z}=\nabla\times{\bf A}$, the mixing can be formulated as ($\hbar, c \equiv 1$)
\begin{eqnarray}
H({\bf k})=
\begin{pmatrix}
H_W({\bf k}-e{\bf A}) & \Delta \\
\Delta^*  &  -\sigma_y H^*_W(-{\bf k}-e{\bf A})\sigma_y
\end{pmatrix}
\label{Hammix}
\end{eqnarray}
with $\Delta=\Delta_0e^{i\phi({\bf r})}$, where $\phi({\bf r})$ is the globally coherent superconducting phase~\cite{SuppleMat}. The Hamiltonian $H_W({\bf k})=v_F {\bf k}\cdot\boldsymbol{\sigma} + \eta_xv_Fk_x\sigma_0 + \beta \sigma_z + \mu \sigma_0$ describes a Weyl node of the WSM layers, where $v_F$ is the Fermi velocity, $\boldsymbol{\sigma}$ is the vector of Pauli matrices and $\eta_x$ indicates the tilt of Weyl cones.

The phase factors $e^{i\phi}$ in the pair potential can be removed from the off-diagonal blocks of Eq.~(\ref{Hammix}) and incorporated into the Weyl Hamiltonian $H_W$ by a gauge transformation~\cite{Anderson.arxiv1998} $H \to \mathcal{U}^\dagger H \mathcal{U}$, where
\begin{eqnarray}
\mathcal{U} =
\begin{pmatrix}
e^{i\phi({\bf r})} & 0 \\
0 & 1
\end{pmatrix}.
\label{AnderGT}
\end{eqnarray}
Then, following Refs~\onlinecite{PBaireuther.NJP2017, Pacholski.PRL2018}, the transformed Hamiltonian can be block diagonalized using a unitary transformation $\mathcal{V}=\exp{(\tfrac{1}{2}i\vartheta \nu_y \sigma_z)}$, with $\tan \vartheta = -\tfrac{\Delta_0}{v_F k_z}$, $\vartheta \in (0, \pi)$, and $\boldsymbol{\nu}$ being the vector of Pauli matrices that act on the electron-hole index. Finally, the two low-energy QP bands are described by (setting $\mu=0$)~\cite{HamiltSI}
\begin{eqnarray}
\begin{aligned}
\mathcal{H}_{BW} =& v_F \sum_{\alpha=x,y}(k_\alpha+a_\alpha + \kappa mv_{s,\alpha})\sigma_\alpha + (\beta-\Gamma_{k_z})\sigma_z   \\
&+ \eta_xv_F(k_x+a_x + \kappa mv_{s,x})\sigma_0,
\end{aligned}
\label{HamHWeyl}
\end{eqnarray}
where $\Gamma_{k_z}=\sqrt{\Delta_0^2 + v_F^2k_z^2}$ and $\kappa=-v_Fk_z/\Gamma_{k_z}$. Here, we have defined a gauge field ${\bf a}= \tfrac{1}{2}\nabla \phi $ and the supercurrent velocity $\boldsymbol{v}_s =(\tfrac{1}{2}\nabla \phi - e{\bf A})/m$. Significantly, when $v_F^2k_z^2=\beta^2-\Delta_0^2$, $\mathcal{H}_{BW}$ respects a generalized chiral symmetry: $\zeta^\dagger\mathcal{H}_{BW}\zeta=-\mathcal{H}_{BW}$ with $\zeta=(\sigma_z-i\eta_x\sigma_y)/\sqrt{1-\eta_x^2}$, which results in a completely dispersionless zeroth LL~\cite{KawarabayashiPRB2011,KawarabayashiIJMP2012}.

The Hamiltonian Eq.~(\ref{HamHWeyl}) describes a pair of  quasielectron BW cones within the effective vector potential $e\boldsymbol{\mathcal{A}}={\bf a} + \kappa m\boldsymbol{v}_s$, which has been shown to produce Landau quantization of Bogoliubov particles~\cite{Pacholski.PRL2018}. The other pair of  quasihole BW cones with an effective vector potential $e\boldsymbol{\mathcal{A}}={\bf a} - \kappa m\boldsymbol{v}_s$ can be obtained by mixing $H_W(-{\bf k})$ and the corresponding hole cone, similar to what is executed in Eq.~(\ref{Hammix}). The low-energy excitations in the heterostructure Weyl superconductor are dominated by normal-state Hamiltonian. The pair potential $\Delta({\bf r})$ only introduces a superconducting gauge field and alters the positions of the nodes. Consequently, the supercurrent velocity $\boldsymbol{v}_s$ always couples to the BW fermions as a vector potential and does not scatter the LLs.

Intrinsic Weyl superconductivity has been proposed in stoichiometric materials such as \ce{URu_2Si_2}~\cite{KasaharaPRL2007,Goswami.arxiv2013}, \ce{UPt_3}~\cite{TouHPRL1998,GoswamiPRB2015}, \ce{UCoGe}~\cite{HuyNTPRL2007}, and \ce{SrPtAs}~\cite{BiswasPKPRB2013,FischerPRB2014}. For example, a Weyl superconducting phase in \ce{URu_2Si_2} may be realized via a chiral $d$-wave paired state $\Delta({\bf k}) = \Delta k_z(k_x+ik_y)/k_F^2$~\cite{KasaharaNJP2009}, with an extra nodal ring in the $k_z=0$ plane, where $k_F$ is the Fermi momentum. To focus on the effect of Weyl points, we reduce the model to a $k_x+ik_y$ paired state. In a magnetic field, the intrinsic Weyl superconductivity can be captured by
\begin{eqnarray}
H_p=
\begin{pmatrix}
\hat{H_e} & \hat{\Delta}  \\
\hat{\Delta}^* & - \hat{H^*_e}
\end{pmatrix},
\label{HamChiralP}
\end{eqnarray}
where the electronic Hamiltonian operator $\hat{H_e} = \tfrac{1}{2m}({\bf k} - e{\bf A})^2 + \eta_xv_F(k_x-eA_x) - \mu$, with ${\bf k}=-i\nabla$. $\hat{\Delta} = -\tfrac{1}{2k_F}\left[\{k_x, \Delta({\bf r})\} + i\{k_y, \Delta({\bf r})\}\right]$ is the gap operator in the continuous limit for the $k_x+ik_y$ paired state, $\Delta({\bf r}) = \Delta$, and curly brackets represent symmetrization, $\{a,b\} 
= ab+ba$.

By performing the Anderson gauge transformation Eq.~(\ref{AnderGT}) on the Hamiltonian $H_p$ and then linearizing the transformed Hamiltonian in the vicinity of a BW node~\cite{Simon.StevenPRL1997,XiangTao2022}, say, $(0,0,k_F)$, we obtain  
\begin{eqnarray}
\begin{aligned}
\mathcal{H}_{BWp}({\bf k}) = &-v_\Delta\sum_{\alpha=x,y} (k_\alpha+a_\alpha) \sigma_\alpha + v_F k_z\sigma_z \\
&+ \eta_xv_F(k_x+a_x)\sigma_0 + \eta_xv_F mv_{s,x}\sigma_z,
\end{aligned}
\label{HamIWeyl}
\end{eqnarray}
where $v_\Delta=\Delta_0/k_F$ is the gap velocity. 

Eq.~(\ref{HamIWeyl}) reveals that, in the mixed state of the intrinsic Weyl superconductor, the effective vector potential coupling to the BW fermions does not include explicitly the external magnetic vector potential. This is apparent because the magnetic vector potential introduced into the BdG Hamiltonian of the superconductor does not appear in the pairing function~\cite{LutchynPRB2009,DiezMPRB2013}. Compared to the result in Eq.~(\ref{HamHWeyl}), the Hamiltonian $\mathcal{H}_{BWp}$ ($k_z=0$) satisfies $\zeta^\dagger\mathcal{H}_{BWp}\zeta=-\mathcal{H}_{BWp}$ only if $v_{s,x}=0$; the supercurrent velocity $\boldsymbol{v}_s$ breaks the generalized chiral symmetry. It is similar to that reported in the mixed state of the single-band $d$-wave superconductor~\cite{Franz.PRL2000,Vafek.PRB63.2001}. 

\begin{figure}[ht]
  \centering
  \subfigure{\includegraphics[width=1.6in]{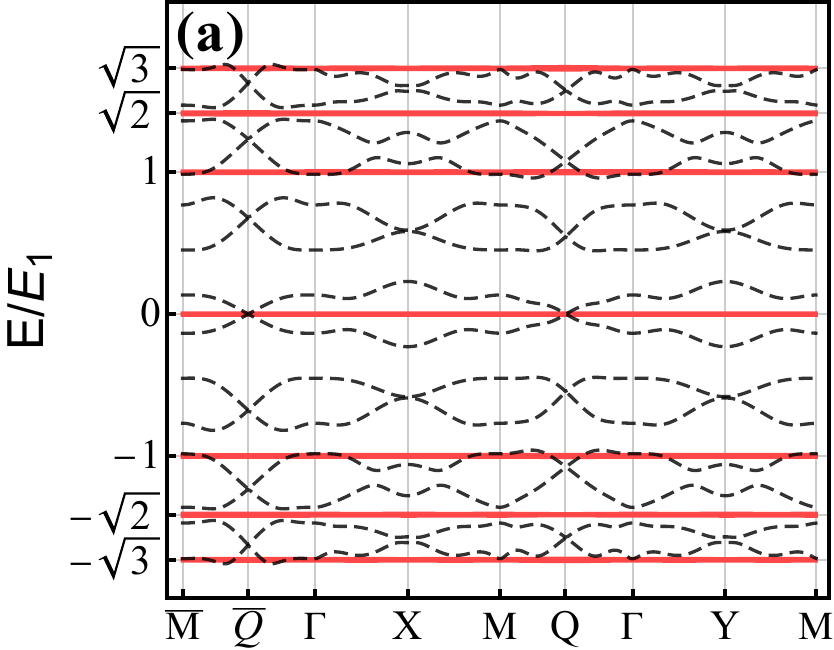}\label{LLa}}~~
  \subfigure{\includegraphics[width=1.6in]{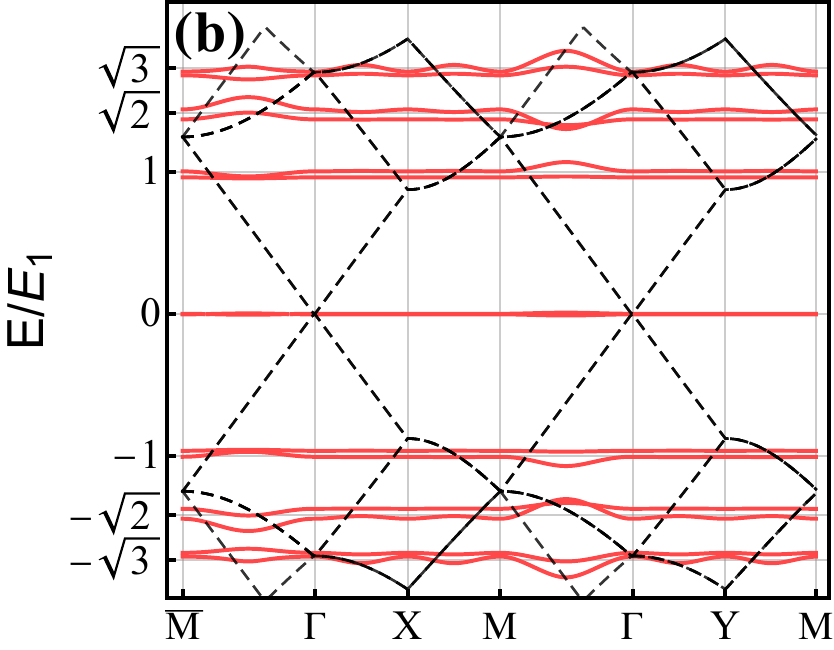}\label{LLb}}
  \caption{(a) The QP spectra in the vortex lattice of two-band (red solid curves) and single-band (black dashed curves) $d$-wave superconductors~\cite{DwaveST}, with $l_B=45$ and $\Delta_0=0.5$. (b) The QP spectra of the 2D heterostructure superconductor in zero magnetic field (black dashed curves) and in the mixed state (red solid curves), with $\beta=\sqrt{2}$, $\mu=0.003$, $l_B=46$ and $\Delta_0=1$. These curves are plotted along the path shown in Fig.~\ref{Lattice}.} \label{LLS}
\end{figure}

\textit{LLs in the mixed state of 2D superconductors.---}
It is ideal to have 2D systems that host LLs where the motion of QPs in the $z$-direction is restricted such that the kinetic energy is quenched and the interaction between QPs dominates. Exotic phenomena such as fractionalized Bogoliubov QPs, the counterpart of fractional electrons in superconductors, may be realized in such systems. The above symmetry analysis in Weyl superconductors can be generalized to 2D superconducting systems. For instance, we demonstrate a Dirac-LL in the mixed state of a two-band $d$-wave superconductor. The BdG Hamiltonian is written as 
\begin{eqnarray}
\mathscr{H}({\bf k})=
\begin{pmatrix}
\mathscr{H}_0({\bf k}) & \Delta ({\bf k}) \\
\Delta^* ({\bf k})  &  -\sigma_y\mathscr{H}^*_0(-{\bf k})\sigma_y
\end{pmatrix},
\label{2ddWSC}
\end{eqnarray}
where the normal-state Hamiltonian $\mathscr{H}_0({\bf k})$ reads 
\begin{eqnarray}
\begin{aligned}
\mathscr{H}_0({\bf k}) = & t_0(\sigma_{y}\sin{k_x} - \sigma_{x}\sin{k_y}) - \mu\sigma_0  + \beta \sigma_z\\
& - t_0\sigma_z(\cos{k_x}+\cos{k_y}),
\end{aligned}
\end{eqnarray}
with the lattice constant $a_0$ set to be 1. The unconventional superconducting phase Eq.~(\ref{2ddWSC}) may be realized by a $d$-wave superconductor heterostructure~\cite{LinderPRL2010,LucignanoPRB2012,WangEryinNP2013}. Considering the $d_{x^2-y^2}$-wave paired state $\Delta({\bf k})=\Delta(\cos{k_x}-\cos{k_y})$. Our numerical calculations (setting $t_0$, $a_0 \equiv 1$) give well-defined Dirac-LLs $E_n=\sqrt{n}E_1$ with $E_1=2\sqrt{\pi} v_F/l_B$ when $\beta=2$, as shown in Fig.~\ref{LLa}. In contrast, for the single-band $d$-wave superconductor, the QPs retain the Dirac cone, and the main effect of the magnetic field is the renormalization of the Fermi velocity~\cite{Yasui.PRL1999,Marinelli.PRB2000,Kopnin.PRB2000,Morita.PRL2001,Knapp.PRB2001,Vishwanath.PRL2001,Vafek.PRL2006,Melikyan.PRB2007}.

We investigate in detail a 2D heterostructure superconductor, of which the BdG Hamiltonian reads
\begin{eqnarray}
\mathscr{H}({\bf k})=
\begin{pmatrix}
\mathscr{H}_0({\bf k}) & \Delta \\
\Delta^*  &  -\sigma_y \tau_0\mathscr{H}^*_0(-{\bf k})\tau_0\sigma_y 
\end{pmatrix},
\label{Ham2DSC}
\end{eqnarray}
with 
\begin{eqnarray}
\begin{aligned}
\mathscr{H}_0({\bf k}) &= \beta\tau_0\sigma_z + t_0\tau_z(\sigma_{x}\sin{k_x}+\sigma_{y}\sin{k_y}) \\
& - \mu\tau_0\sigma_0 + t_0\tau_x\sigma_0(\epsilon-\cos{k_x}-\cos{k_y}),
\end{aligned}
\label{HamTI}
\end{eqnarray}
where $\boldsymbol{\tau}$ is the Pauli matrix that acts on the orbital degree of freedom.

\begin{figure}[ht]
  \centering
  \subfigure{\includegraphics[width=1.7in]{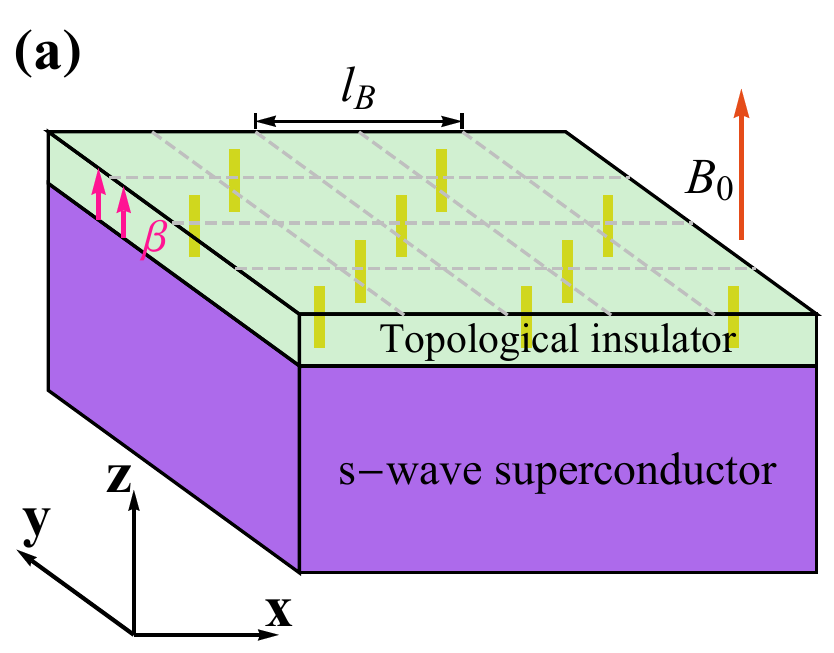}\label{hstrut}}~~
  \subfigure{\includegraphics[width=1.5in]{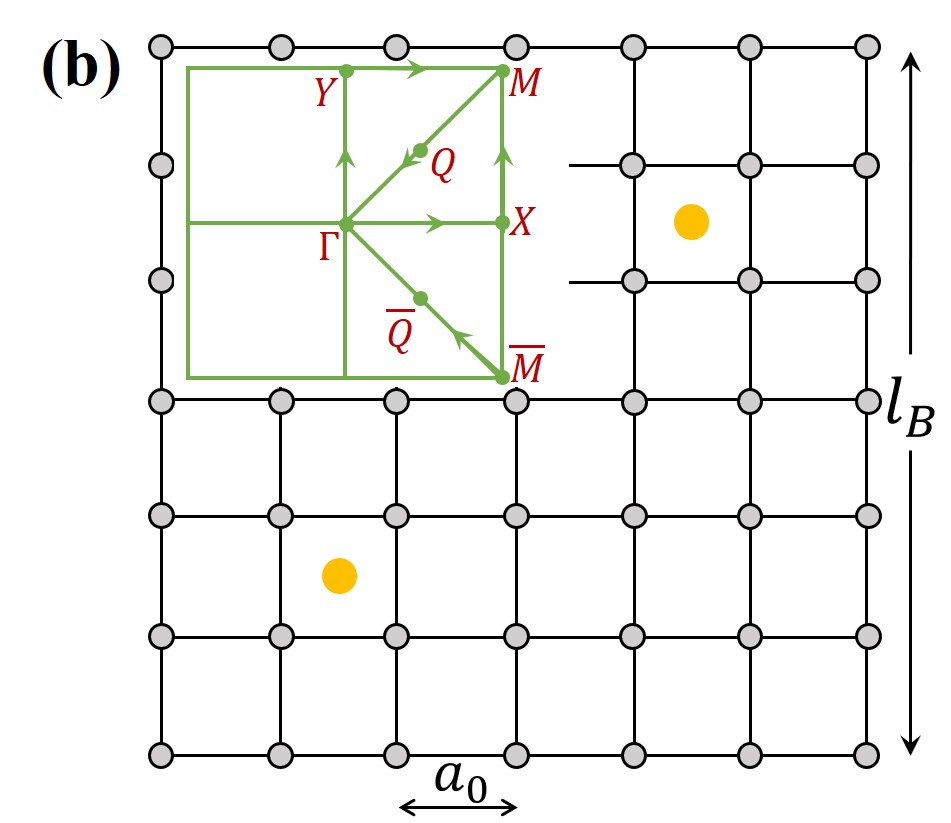}\label{Lattice}}
  \caption{(a) A TI-superconductor heterostructure (a thin layer of a TI, with perpendicular magnetization $\beta$, deposited on a $s$-wave superconductor), in a perpendicular magnetic field $B_0$. (b) Magnetic unit cell (for $l_B=6a_0$) containing two vortices represented by  yellow dots, with each vortex carrying flux $hc/2e$.} \label{hslat}
\end{figure}
The system Eq.~(\ref{Ham2DSC}) can be realized by depositing a thin layer of a TI on a conventional $s$-wave superconductor, as shown in Fig.~\ref{hstrut}. The TI layer described by Hamiltonian Eq.~(\ref{HamTI}) can be obtained from a 3D TI material with layered structures, such as the \ce{Bi_2Se_3} family~\cite{Zhang.NatureP2009}, which is regularized on a simple square lattice~\cite{FuLiangPRL2010,QiXiaoLiang.RMP2011,Vazifeh.PRL2013}.
By tuning the parameter $\epsilon$ (in this work, we set $\epsilon=3$), the 3D TI material can be transformed into a strong TI phase. We note that similar 2D heterostructure superconductors have been extensively studied at both theoretical~\cite{FuLiangPRL2008,FuLiangPRL2009,TanakaPRL2009,SauJayDPRL2010,QiXiao-LiangPRB2010} and experimental~\cite{Mei-XiaoScience2012,XuSu-YangNatureP2014,XuJin-PengPRL2014,XuJin-PengPRL2015,SunHao-HuaPRL2016} levels. On the other hand, our heterostructure model can be viewed as a single layer of the MB model~\cite{Meng.PRB2012,Meng.PRB2017}.

In a zero magnetic field, Hamiltonian Eq.~(\ref{Ham2DSC}) exhibits a Bogoliubov-Dirac node at $(0,0)$ when $\beta^2=1+\Delta_0^2$, which separates two different topological phases. In the presence of a vortex lattice schematically shown in Fig.~\ref{Lattice}, the QP spectrum of the 2D superconductor can be calculated ~\cite{SuppleMat}. A square vortex lattice is formed with a magnetic unit cell $l_B\times l_B$, where the magnetic length $l_B=\sqrt{\phi_0/B_0}$ with $\phi_0=hc/e$ being the flux quantum. The numerically calculated QP spectrum is shown in Fig.~\ref{LLb}, which exhibits a doubly degenerate Dirac-LL structure $E_n=\sqrt{n}E_1$ in the presence of a vortex lattice, with a completely dispersionless zeroth LL. The degeneracy of these Bogoliubov LLs can be lifted by tuning the parameters $\beta$ and $\mu$. For instance, when $\beta$ slightly deviates from the fine-tuned values, the zeroth LL departs from zero energy, and the spectrum exhibits LLs that are similar to those of a massive Dirac fermion with a definite charge in a magnetic field. 

\textit{Magneto-optical conductivity.---}
The magneto-optical conductivity measurement is a widely used method to observe the LL structures in various systems~\cite{VPGusynin.JPCM2007,JiangPRL2007,OrlitaSST2010,Ashby.PRB2013,Zhi-GuoPRL2017}. Here, we investigate the optical excitation of QPs in the mixed state of the 2D heterostructure superconductor. The magneto-optical conductivity tensor in the vortex lattice can be derived from the Kubo formula~\cite{Ahn.NatureC2021,Tianrui.PRB2019,Kamatani.PRB2022,Papaj.PRB2022}. Expressed in the LL basis in the clean limit, we have
\begin{eqnarray}
\sigma_{\alpha \alpha^\prime}(\omega)=- \tfrac{ie^2}{2\pi l_B^2} \sum_{nn^{\prime}} \int \tfrac{d{\bf k}}{(2\pi)^2} \tfrac{f(E_n)-f(E_{n^\prime})}{E_n-E_{n^\prime}} \times \tfrac{V_{\alpha}^{nn^\prime} V_{\alpha^\prime}^{n^\prime n}}{\omega+E_n -E_{n^\prime} + i\delta},
\label{eqMOC}
\end{eqnarray}
where $f(\epsilon)=1/(1+e^{\epsilon/k_BT})$ is the Fermi distribution function and the Boltzmann constant $k_B \equiv 1$ in our numerical calculations, $\alpha=\{x,y\}$, and $d{\bf k}=dk_xdk_y$. The velocity matrix element $V_{\alpha}^{nn^\prime} = \langle \psi_{n{\bf k}} |V_{\alpha}| \psi_{n^\prime{\bf k}} \rangle$ with $| \psi_{n{\bf k}} \rangle$ being the eigenvector of the $n$-th LL, satisfying $\mathcal{H}_M|\psi_{n{\bf k}} \rangle = E_n |\psi_{n{\bf k}} \rangle$ ($\mathcal{H}_M$ can be found in the Supplemental Material~\cite{SuppleMat}). The velocity operator in the superconducting state is given by~\cite{Ahn.NatureC2021}
\begin{eqnarray}
V_{\alpha} ({\bf k}) = 
\begin{pmatrix}
\partial_{k_\alpha} H^{\bf B}_0({\bf k}-e{\bf A}) & 0 \\
0 & \partial_{k_\alpha} H_0^{{\bf B}*}(- {\bf k}-e{\bf A})
\end{pmatrix},
\end{eqnarray}
where $H^{\bf B}_0$ can be obtained from the BdG Hamiltonian $\mathcal{H}_M$ of the 2D heterostructure superconductor.

\begin{figure}[ht]
  \centering
  \subfigure{\includegraphics[width=2.2in]{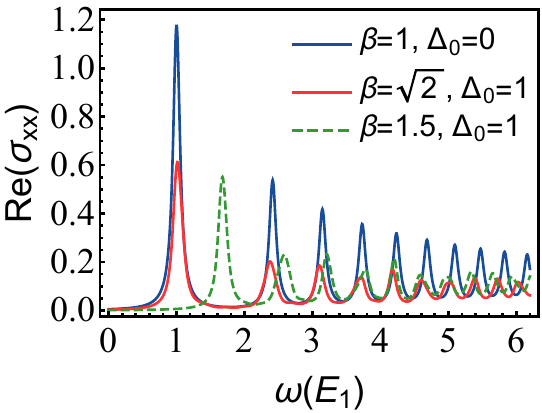}\label{MOC}}
  \caption{The real part of the magneto-optical conductivity (in units of $e^2/2\pi l_B^2$) of the 2D heterostructure superconductor, with $L=22$, $\mu=0$, $T=0.001$ and the dimensionless $\delta=0.01$. $\Delta_0=0$ indicates a Dirac-semimetal phase. The red and blue solid curves conform to the gapless condition $\beta^2 = 1+\Delta_0^2$.} \label{MOCond}
\end{figure}

The real part of the calculated magneto-optical conductivity of the 2D superconductor is shown in Fig.~\ref{MOCond}. The blue solid curve represents the results of the Dirac-semimetal phase ($\Delta_0=0$) and gives a graphene-like optical conductivity, which exhibits a series of peaks at $\omega/E_1 = \sqrt{n}+\sqrt{n+1}$ ($n\ge 0$) because the major contribution to the conductivity comes from the $n\to n\pm 1$ transitions~\cite{VPGusynin.JPCM2007}. In the superconducting phase, as demonstrated by the red solid curve, the magneto-optical conductivity due to the Bogoliubov LLs is similar to that in the Dirac-semimetal phase, but with lower amplitudes. Moreover, when $\beta_2 \ne 1+\Delta_0^2$, the Bogoliubov LLs deviate from $\sqrt{n}E_1$, causing the conductivity peaks to shift from $\sqrt{n}+\sqrt{n+1}$, as shown by the green dashed line.

\textit{Quantum thermal Hall effect.---}
We also numerically calculate the Chern numbers of the (non-degenerate) Bogoliubov LLs of the 2D superconductor Eq.~(\ref{Ham2DSC}), yielding $1$ (or $-1$) for each LL, which is the same as that of the electric LLs in the normal state~\cite{BAndrei2013}. Hence, as demonstrated in the following, the superconducting state and the normal state can exhibit similar transverse conductivity tensors. While a 2D Landau quantized electron gas can show a quantized Hall conductivity $\sigma_{xy}=Ne^2/h$ when the Fermi level lies in a gap between two adjacent LLs, where $N$ denotes the sum of the Chern numbers over all occupied LLs~\cite{ThoulessDJPRL1982}, Bogoliubov QPs do not carry a definite charge and the electric Hall conductivity is not expected to be quantized. Instead, we expect a quantized thermal Hall response resulting from these Bogoliubov LLs in the 2D superconductor. 

We investigate the thermal Hall conductivity of the 2D heterostructure superconductor. In the mixed state, the thermal transport coefficient $\kappa_{xy}$ is given by~\cite{Cvetkovic.NatureC2015,Vafek.PRB64.2001}
\begin{eqnarray}
\kappa_{xy} = \tfrac{1}{\hbar T} \int d{\xi} \xi^2 \left(-\tfrac{\partial f(\xi)}{\partial \xi} \right) \tilde{\sigma}_{xy}(\xi)
\label{ThermalC},
\end{eqnarray}
where $f(\xi)=1/(1+e^{\xi/k_BT})$. The transverse conductivity tensor $\tilde{\sigma}_{xy}$ is defined as
\begin{eqnarray}
\tilde{\sigma}_{xy}(\xi) = -\tfrac{i}{2\pi} \int d{\bf k} \sum_{E_{n^\prime} ({\bf k}) < \xi < E_n({\bf k})} \tfrac{J^x_{n^\prime n}({\bf k}) J^y_{n n^\prime} ({\bf k}) - J^y_{n^\prime n}({\bf k}) J^x_{n n^\prime} ({\bf k})}{(E_n({\bf k}) - E_{n^\prime} ({\bf k}))^2},
\end{eqnarray}
with $J^{\alpha}_{n^\prime n}({\bf k}) = \langle n^\prime {\bf k}| \tfrac{\partial \mathcal{H}_M}{\partial k_{\alpha}}|n{\bf k} \rangle$. It can be expressed as $\tilde{\sigma}_{xy}(\xi) = \sum_n\Omega_n(\xi)$, where $\Omega_n(\xi)$ represents the integral of the Berry curvature of the $n$th LL over the region of the magnetic Brillouin zone that lie below the QP energy $\xi$. If the LL is fully occupied, then the integral of the Berry curvature is the Chern number. The tensor $\tilde{\sigma}_{xy}(\xi)$ has a simple physical interpretation: it is proportional to the contribution of QPs to the $T=0$ spin Hall conductivity when all the LLs below the energy $\xi$ are occupied~\cite{Cvetkovic.NatureC2015}. Additionally, when $\Delta_0=0$, the conductivity tensor $\tilde{\sigma}_{xy}(\xi)$ is proportional to the usual electric Hall conductivity.
 
\begin{figure}[ht]
  \centering
  \subfigure{\includegraphics[width=1.55in]{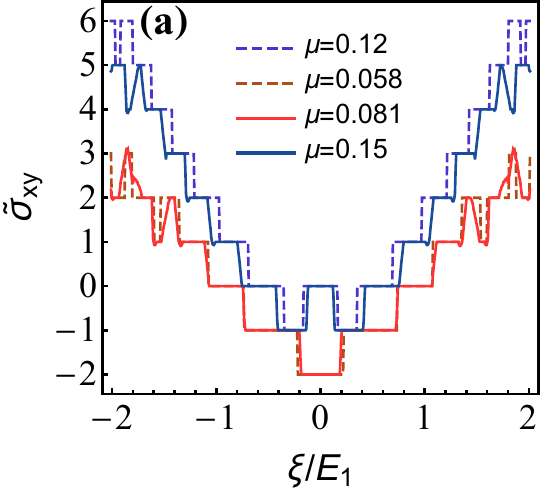}\label{HallC}}~~
  \subfigure{\includegraphics[width=1.7in]{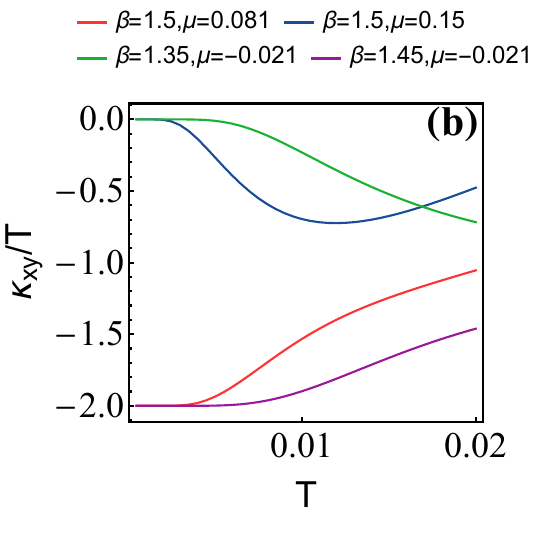}\label{ThermS}}
  \caption{(a) The transverse conductivity tensor $\tilde{\sigma}_{xy}$ of the 2D superconductor as a function of the QP energy $\xi$, with $l_B=22$. The solid (dashed) lines denote the superconducting phase (normal state), with $\beta=1.5$ and $\Delta_0=1$ ($\beta=1.1$ and $\Delta_0=0$). (b) The thermal Hall conductivity $\kappa_{xy}/T$ (in units of $\pi^2/3$, $k_B\equiv1$), expressed as a function of temperature $T$, with $l_B=22$ and $\Delta_0=1$~\cite{ThermalHC}.} \label{Hall_Ther}
\end{figure}
In Fig.~\ref{HallC}, we show the numerically calculated transverse conductivity tensor $\tilde{\sigma}_{xy}(\xi)$ for the superconducting state ($\Delta_0=1$) and the normal state ($\Delta_0=0$), respectively. Our calculations show that in the vortex state of the 2D heterostructure superconductor, the transverse conductivity tensor exhibits a series of plateaus by tuning the parameters $\beta$ and $\mu$, with each plateau corresponding to $\xi$  moving through a gap between two adjacent Bogoliubov LLs. In the low temperature limit, as illustrated in Ref.~\onlinecite{Vafek.PRB64.2001}, the thermal transport coefficient $\kappa_{xy}$ in Eq.~(\ref{ThermalC}) approaches 
\begin{eqnarray}
\kappa_{xy} = \tfrac{\pi^2}{3} \left( \tfrac{k_B}{\hbar} \right)^2 T \tilde{\sigma}_{xy}(0).
\end{eqnarray}
Consequently, when $T\to 0$, the ratio $\kappa_{xy}/T$ is quantized due to the quantization of the transverse conductivity tensor $\tilde{\sigma}_{xy}$, as shown in Fig.~\ref{ThermS}.

\textit{Discussion and conclusion.---}
The 2D superconductor described by Hamiltonian Eq.~(\ref{Ham2DSC}) can exhibit a Bogoliubov Fermi surface by tuning the chemical potential $\mu$ of the TI layer. Quantum oscillations induced by such a Fermi surface in a strained Weyl superconductor have been investigated~\cite{LiuTianyuPRB2017}. Therefore, a multi-plateau quantum thermal Hall effect may be observed in the mixed state of the 2D heterostructure superconductor. Additionally, the pseudo-LLs of Bogoliubov particles, a LL-like spectrum induced by a pseudovector potential, have also been reported in several gapless superconducting systems when applying a pseudomagnetic field produced by lattice strain~\cite{LiuTianyuPRB2017,Massarelli.PRB2017,EmilianMPRB2018,Guinea.NatureP2010,Levy.Science2010}. The Bogoliubov pseudo-LLs can also lead to quantum thermal Hall effect.

A further question is whether Bogoliubov QPs in the LLs of the 2D superconductor Eq.~(\ref{Ham2DSC}) can be fractionalized when interactions between the QPs are included.  The distribution of the Berry curvature is an essential factor. The normal state LLs feature constant Berry curvature that is proportional to $\tfrac{1}{B_0}$~\cite{OzawaPRB2021}; in contrast, the Berry curvature of the Bogoliubov LLs in the superconducting phase is strongly momentum-dependent due to the presence of the internal gauge field ${\bf a}$, except for the zeroth LL, which exhibits almost uniform Berry curvature~\cite{SuppleMat}.  Consequently, fractional quantum thermal Hall effect may be realized in the zeroth Bogoliubov LL, as it manifests as a topologically ideal flat-band~\cite{WuYang-LePRB2012,RoyPRB2014,WangJiePRR2023}.

In summary, we have investigated several nodal superconducting systems and identified a crucial ingredient for the LL quantization of Bogoliubov particles: the low-energy excitations need to be dominated by the normal-state physics, which has often been neglected in previous studies. We proposed a 2D heterostructure superconductor that can exhibit Bogoliubov LLs in the presence of a vortex lattice. In particular, such LL structures may lead to peaked magneto-optical conductivity and a quantized thermal Hall conductivity $\kappa_{xy}/T$ in the mixed state of the 2D superconductor in the low-temperature limit.

\textit{Acknowledgements.---}
This work was supported by Guangdong Basic and Applied Basic Research Foundation (Grant No. 2021B1515130007), the National Natural Science Foundation of China (Grant No. 12004442), and Shenzhen Natural Science Fund (Grants No. 20231120172734001 and 20220810130956001).

\normalem
\bibliography{LLs_in_2DSC}

\begin{thebibliography}{79}
\bibitem [{Ham()}]{HamiltSI}%
  \BibitemOpen
  \href@noop {} {}\bibinfo {note} {Hamiltonian~(\ref{HamHWeyl}) does not include virtual transitions to higher bands, of second order in $v_s$.}\BibitemShut {Stop}%
\bibitem [{The()}]{ThermalHC}%
  \BibitemOpen
  \href@noop {} {}\bibinfo {note} {The values of $\tilde{\sigma}_{xy}(0)|_{\beta=1.5, \mu=0.081}$ and $\tilde{\sigma}_{xy}(0)|_{\beta=1.5, \mu=0.15}$ are shown in Fig.~\ref{HallC}, and our calculations also yield $\tilde{\sigma}_{xy}(0)|_{\beta=1.35, \mu=-0.021}=0$ and $\tilde{\sigma}_{xy}(0)|_{\beta=1.45, \mu=-0.021}=-2$.}\BibitemShut {Stop}%
\end{thebibliography}

\end{document}